# THE IMPACT OF INTEREST RATES ON FIRM'S FINANCIAL DECISIONS

**Efendi[1*], Rahmadani Srifitri[1], Septriza Berliana[1]**

Department of Mathematics and Data Science[1],
Faculty of Mathematics and Natural Sciences, Universitas Andalas
Limau Manis, Kec. Pauh, Padang, West Sumatra, 25166

[*]E-mail: efendi@sci.unand.ac.id

**Abstract.** Financial decisions are the decisions that managers take with regard to the finances of a company. This article aims to examine and explain the effect of interest rates on economic and financial decisions such as investment, funding, and dividend in a firm. This research uses the correlation coefficient analysis methods and descriptive methods to illustrate the relationship between interest rates and financial decisions. The data used in this research was obtained from several government reports and leading economic sources. The results of this research show that interest rates have a negatively insignificant effect on investment and funding decisions, but positively moderate effect on dividend decisions.

## 1. INTRODUCTION

Attaining the highest company value is possible by implementing financial management functions, wherein each financial decision made not only influences other financial choices but also has a direct impact on the overall value of the company. In the pursuit of company objectives, it becomes the duty and responsibility of company managers to make judicious decisions and formulate appropriate policies (Mulyanti, 2017). Financial decisions encompass the management of financial resources derived from diverse sources, and the determination of the type of funding source, financing period, financing costs, and expected returns becomes crucial.

Furthermore, the dynamic nature of the business environment requires financial managers to adapt and respond swiftly to changes in economic conditions, market trends, and regulatory landscapes. In this context, staying abreast of industry developments and employing a forward-looking approach in financial decision-making becomes imperative. Financial managers must assess risks meticulously, considering factors such as market volatility and economic uncertainties, to devise strategies that not only optimize company value but also safeguard against potential setbacks. Effective communication and collaboration with other departments within the organization are also crucial, as financial decisions often have cross-functional implications. By fostering a comprehensive understanding of the intricate interplay between financial choices and broader business objectives, financial managers contribute significantly to the sustained success and resilience of the company in an ever-evolving business landscape. Below are some responsibilities for financial manager (Nurmasari & Nur'aidawati, 2021).

1. Take active investment/spending decisions (investment decisions)
   Selecting the desired investment from several available opportunities, deciding on one or more investment alternatives that are considered the most profitable [1].
2. Making passive funding/spending decisions (financing decisions)
   Choose from various sources of funds available to invest, deciding on one or more funding alternatives that provide the most efficient costs.
3. Taking dividend decisions

Relating to the issue of determining the percentage of profits to be paid as cash dividends to shareholders, consistency in dividend payments, issuance of dividend shares, and share repurchase.

The interest rate is one of external element that can influence a company's financial decision making. Interest rate is the ratio between interest and the amount of money borrowed or invested for a certain period of time (Fahlevi, 2019). In the banking sector, transactions cannot be avoided from the significant influence of interest rates. The amount of the loan is called the principal or principal value, while the percentage of the principal that must be paid as fees or interest within a certain period is known as the interest rate (Fahlevi, 2019).

If interest rates increase unreasonably, the business world will face difficulties in paying interest expenses and obligations. High interest rates will increase the burden on the company, resulting in a direct reduction in company profits. Overall, a reduction in interest rates will encourage economic growth because the flow of funds will increase. Therefore, interest rate and profit factors have an important role that greatly influences financial decision making. But, actually, Karpavicius and Yu in (Sigitas Karpavičius a, 2017) had shown the impact of interest rates on firm's financing policies, that slightly negative on US industrial firms. So, they suggested that firms do not adjust their capital structures based on interest rates, except that real gross domestic product will be negative. With an increase in interest rates, businesses with company credit cards and existing loans can have higher interest payments, less disposable income and bigger overheads. In some cases, borrowers may find themselves paying off the interest only, rather than the loan itself. In this article, we will analyze how the interests rates impacts on P.T. Astra International Tbk, an industry in Indonesia,

## 2. PRELIMINARIES

**Interest Rates**

According to (Prasetyo & Firdaus, 2009), interest is the value of funds that can be borrowed or invested. In classical theory, the term interest is considered to be the price that occurs in the investment market. The investment itself is also influenced by interest rates. In this perspective, interest rates are the cost of capital for the company.

When a company plans to meet its capital needs, this decision is greatly influenced by the interest rate prevailing at that time. This interest rate will determine the type of capital that will be used by the company. For example, whether the company will issue bonds or issue shares. Issuance of shares is chosen if the current interest rate is lower than the level of earning power from the additional capital that will be obtained (Riyanto, 2001).

**Investment Decision**

Investment is the action of releasing assets currently owned with the aim to obtain assets in the future with a higher value. Meanwhile, results in China for 1997-2006 found that investment in fixed assets, government spending on education and health as a proxy for human capital, and infrastructure development positively affected regional convergence in investment decisions are represented by the Price Earnings Ratio (PER), which is the ratio between the stock market price and earnings per share. The use of PER in this research was chosen because it reflects the market's appreciation of the company's ability to generate profits.

**Funding Decisions**

Funding decision is defined as the choice of funding composition made by the company. This decision can be measured using the Debt to Equity Ratio (DER). Debt to Equity Ratio (DER) is a ratio that shows the comparison between funding through debt and funding through equity

(Birmingham, E.F. & Houston, 2001). The Indonesian Stock Exchange (2008) explains that DER is often referred to as the leverage ratio because it reflects the company's capital structure.

**Dividend Decision**

Dividend policy is a decision to determine how much or what proportion of profits will be distributed as dividends to shareholders. Dividend policy is represented using the Dividend Payout Ratio (DPR), indicates how large a share of dividends is taken from the company's net profit.

3. **RESEARCH METHODS**

   **Data Collection**

   P.T. Astra International Tbk was the sample in this research. The data consists of investment decisions, funding decisions, and dividend decisions which are secondary data obtained from the IDX website (Indonesian Stock Exchange).

   **Operational Definition of Research Variables**

   1. Interest Rate:

      Interest rates are the cost of capital for companies. The data uses data sourced from *https://tradingeconomics.com/*, that shows BI rate interest rate per year in period 2018-2022 in the form of a percentage (%).

   2. Investment Decision

      Investment decisions involve a combination of currently owned assets (assets in place) and future investment choices that have a positive present value (net present value). The internal rate of return (IRR) indicator cannot be observed directly (latent), so the calculation uses the Price Earnings Ratio (PER) is formulated as:

$$PER = \frac{Share\ Price}{EPS}$$

Where: EPS = Earning of Pay-out of Shares

3. Funding Decisions

    Funding decisions refer to choices regarding the composition of funding sources made by a company. In this research, funding decisions are confirmed through Debt to Equity Ratio (DER) analysis defined by:

$$DER = \frac{Total\ Debt}{Total\ Equity}$$

4. Dividend Decision

    Dividend policy decisions are decisions regarding the amount of current profits that was be distributed as dividends, compared with the amount of profits that was retained for reinvestment in the company (Birmingham, E.F. & Houston, 2001). In this research, this dividend policy/decision is represented by the Dividend Payout Ratio (DPR) is formulated by equation below:

$$DPR = \frac{DPS}{EPS}$$

Where:  DPS = Dividend Per Shares
EPS = Earning Per Shares

**Data Organizing**

The data was analyzed and arranged in such a way as to provide a structured understanding of the relationship between interest rates and company financial decisions. The data was arranged in tabular form to simplify the interpretation and analysis process.

**Descriptive Analysis**

A descriptive analysis approach were applied to analyze the data that has been collected in detail. This approach involves identifying patterns, trends, and characteristics related to the relationship between interest rates and corporate financial decisions. Data were interpreted using descriptive statistical methods such as average, median, and percentile to describe the overall data distribution. The analysis allowed for a comprehensive examination of the data. This approach involves a meticulous exploration of patterns, trends, and characteristics that pertain to the intricate interplay between interest rates and corporate financial decisions. To offer a detailed overview of the data distribution, various descriptive statistical methods, including measures such as average, median, and deviation, are utilized in the interpretation process.

**Correlation coefficient**

Correlation coefficient analysis is used to identify the direction and strength of the relationship between two or more variables. The direction of the relationship is expressed as positive or negative, while the strength of the relationship is expressed based on the correlation coefficient value (Sugiyono, 2018). The correlation coefficient analysis is employed to discern the direction and strength of relationships among multiple variables. This analytical tool facilitates the expression of the relationship direction as either positive or negative, while the correlation coefficient value quantifies the strength of this relationship. By integrating this quantitative aspect, the study aims to provide a more robust and precise understanding of the intricate dynamics between interest rates and various facets of corporate financial decision-making.

The correlation coefficient $(r)$ functions to show the degree of correlation between the independent variable and the dependent variable. The correlation coefficient value must be within the limits of $-1$ to $+1$ $(-1 < r \leq +1)$, which produce several probabilities, as follows:

a. A positive sign means that there is a positive correlation in the variables being tested, which means that every increase and decrease in the values of $X$ will be recognized by an increase and decrease in $Y$. If $r = +1$ or close to 1, then there is a positive influence between the variables being tested very strong (Sugiyono, 2018).

b. A negative sign means that there is a negative correlation between the variables being tested, meaning that every increase in the $X$ values will be followed by a decrease in the $Y$ values and vice versa. If $r = -1$ or close to -1, then it shows that there is a negative influence and the correlation of the variables being tested is strong (Sugiyono, 2018).

c. If $r = 0$ or close to 0, then the correlation is weak or there is no correlation at all between the variables studied and tested (Sugiyono, 2018).

Following the completion of both descriptive analysis and correlation coefficient analysis, the study synthesizes key insights concerning the correlation between interest rates and corporate financial decisions. The conclusions drawn encapsulate significant discoveries derived from the analytical processes, shedding light on their implications for the realm of corporate financial decision-making.

5. **RESULT AND DISCUSSION**

The financial data of PT. Astra International Tbk. and annual interest rate in period 2018-2022 are shown in Table 1 and Figure 1. To find the impact of interest rate on PT. Astra International Tbk financial policies we used both descriptive analysis and correlation coefficient analysis. We also compare to another research finding in US industrial firms on the impact of interest rates on firm's financing policies, that slightly negative as stated by (Sigitas Karpavičius a, 2017).

**Table 1**
**Astra Company Investment Decisions, Funding Decisions and Dividend Decisions, and Annual Interest Rates, 2018-2022**
*Source : www.tradingview.com and https://tradingeconomics.com/*

| Year | Investment Decision (PER) | Funding Decision (DER) | Dividend Decision (DPR) | Interest Rates (IR) |
|---|---|---|---|---|
| 2018 | 15.36 | 0.63 | 0.400243 | 4.25 |
| 2019 | 14.6 | 0.62 | 0.399254 | 6 |
| 2020 | 13.06 | 0.51 | 0.285714 | 4.75 |
| 2021 | 12.19 | 0.42 | 0.478958 | 4.25 |
| 2022 | 11.64 | 0.37 | 0.895105 | 6 |
| Total | 66.85 | 2.55 | 2.459274 | 25.25 |
| **Transformed Data** | | | | |
| 2018 | 0.229768 | 0.247059 | 0.162748 | 0.168317 |
| 2019 | 0.218399 | 0.243137 | 0.162346 | 0.237624 |
| 2020 | 0.195363 | 0.2 | 0.116178 | 0.188119 |
| 2021 | 0.182349 | 0.164706 | 0.194756 | 0.168317 |
| 2022 | 0.174121 | 0.145098 | 0.363971 | 0.237624 |
| **Descriptive** | | | | |
| Average | 0.2 | 0.2 | 0.2 | 0.2 |
| Median | 0.195363 | 0.2 | 0.162748 | 0.188119 |
| Deviation | 0.002227 | 0.008335 | 0.036746 | 0.00498 |

In Table 1, data was transformed to percentage to provide an understanding structure of the comparison between interest rates and company financial decisions. The average and median of PER and DER are similar to IR. It means the central tendency of PER and DER are similar to IR, but the central tendency of DPR are not similar. So that, descriptively we have found that from 2018-2022, the influence of interest rate to the Investment Decision and Funding Decision of PT. Astra International Tbk are quiet similar, but the influence of interest rate to Dividend Decision are not similar. To find how the interest rate impact on firm's financial policies trend, we check linear trend as shown in Figure 1.

In Figure 1, data was presented in bar chart to simplify the interpretation and analysis of visual trend. Visually, the chart shown that PER and DER were decreased from year to year, but

DPR shown increased significantly. Descriptively, we can conclude that interest rate had negatively significant impact on Investment Decision and Funding Decision of PT. Astra International Tbk. On the other hand, we conjecture that interest rate had positively significant impact on Dividend Decision, but we need another analysis, that is correlation coefficient analysis to find how the data correlate each other.

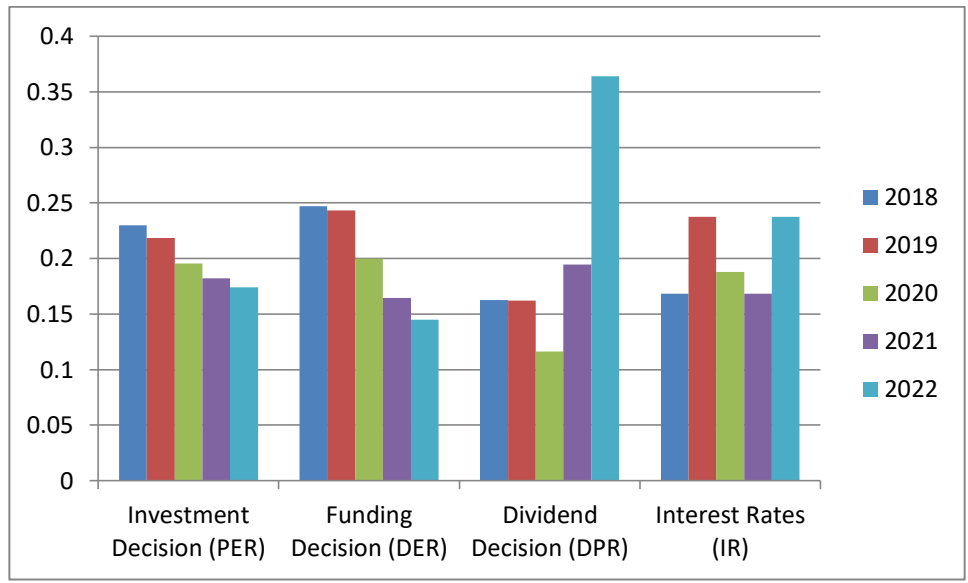

**Figure 1.** The bar chart of PER, DER, DPR, IR from 2020 to 2022

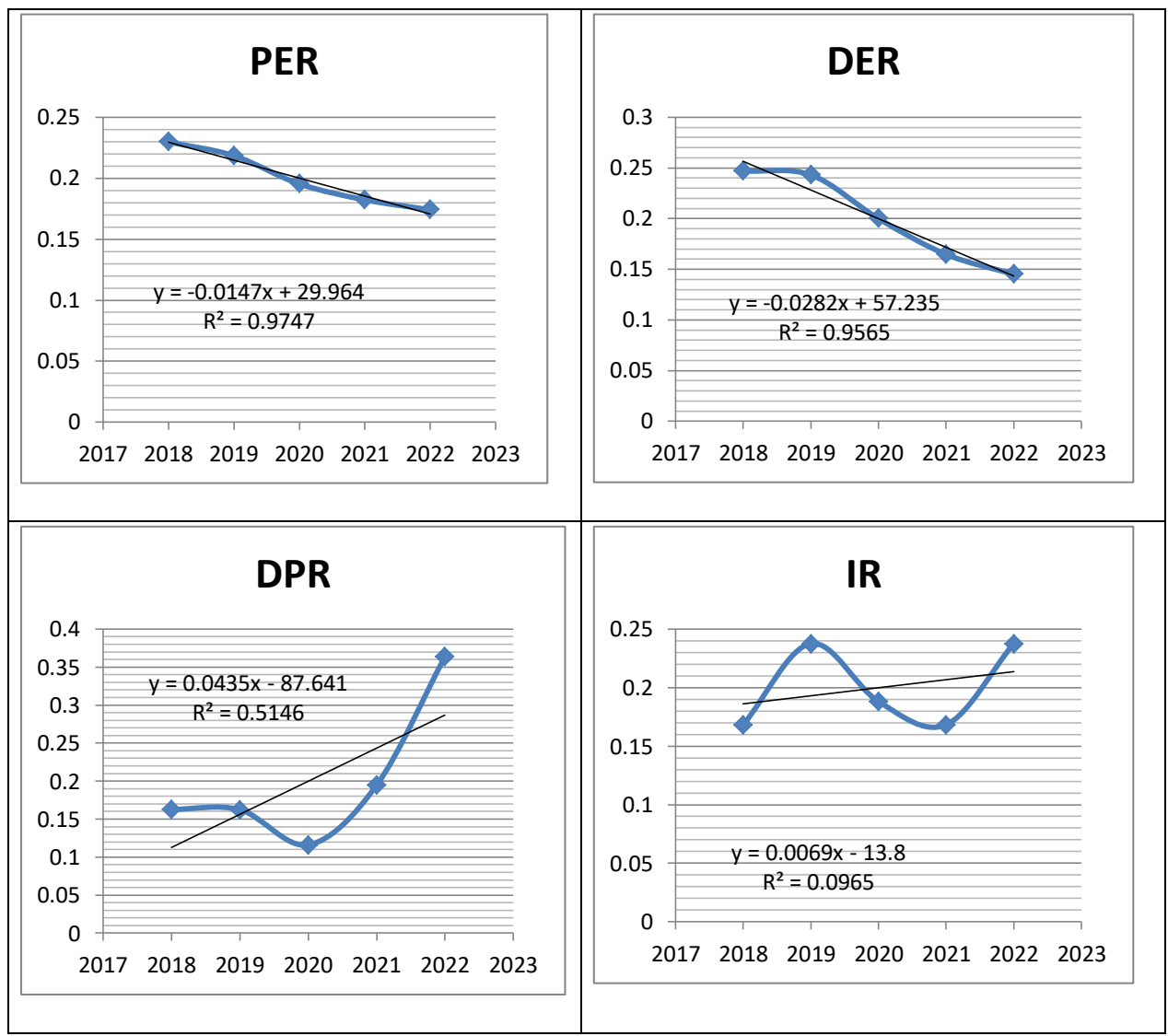

**Figure 2.** Linear trend of PER, DER, DPR, IR from 2018 to 2022

Meanwhile, as a visual linear trend shown in Figure 2, we can see the coefficient of linear trend of PER is -0.0147 and DER is -0.0282, while IR is +0.0069 and DPR +0.0435, we can conclude that interest rate had negatively significant impact on Investment Decision and Funding Decision of PT. Astra International Tbk, but positively significant impact on Dividend Decision.

**Correlation Coefficient Analysis**

The correlation coefficient between the Interest Rate (IR) and each company's financial decisions can be seen in Table 2, as follows:

**Table 2.** The correlation coefficient between the Interest Rate

| Year | Investment Decision (PER) | Funding Decision (DER) | Dividend Decision (DPR) | Interest Rates (IR) |
|---|---|---|---|---|
| 2018 | 15.36 | 0.63 | 0.400243 | 4.25 |
| 2019 | 14.6 | 0.62 | 0.399254 | 6 |
| 2020 | 13.06 | 0.51 | 0.285714 | 4.75 |
| 2021 | 12.19 | 0.42 | 0.478958 | 4.25 |
| 2022 | 11.64 | 0.37 | 0.895105 | 6 |
| Total | 66.85 | 2.55 | 2.459274 | 25.25 |
| Correlation to interest rate | -0.18321 | -0.12656 | 0.52448 | 1 |

**The Effect of Interest Rates on Investment Decisions (PER)**

From Table 2, the correlation coefficient between Interest Rates (IR) and Investment Decisions (PER) obtained is -0.18321. Thus, the two variables have a weak negative correlation where the higher the interest rate, the lower the investment decision, and vice versa. However, because the correlation is weak, it can be said that interest rates do not significantly influence a company's Investment Decisions.

From the descriptive analysis, it appears that changes in interest rates have an influence on company investment decisions. When interest rates decrease, companies tend to be more interested in making new investments or expanding existing investments. This is due to lower funding costs and the availability of greater financial resources. However, conversely, when interest rates rise, a company's investment decisions may become more conservative.

**Impact of Interest Rates on Funding Decisions (DER)**

From Table 2, the correlation coefficient between Interest Rates (IR) and Funding Decisions (DER) obtained is -0.12656. Thus, the two variables have a weak negative correlation where the higher the interest rate, the lower the funding decision, and vice versa. However, because the correlation is weak, it can be said that interest rates do not significantly influence a company's Funding Decisions.

The results of descriptive analysis show that interest rates influence company funding decisions. When interest rates are low, companies tend to seek external funding sources such as bank loans or bonds to fund new projects. Conversely, when interest rates are high, companies may prefer to use internal funding sources or reduce debt levels in order to reduce funding costs.

**Relationship between Interest Rates and Dividend Decisions (DPR)**

The correlation coefficient between Interest Rates (IR) and Dividend Decisions (DPR) obtained is 0.52448. Thus, the two variables have a moderate positive correlation where the higher the interest rate, the higher the dividend decision, and vice versa. However, because the correlation is moderate, it can be said that interest rates moderately influence a company's Dividend Decisions.

Furthermore, descriptive analysis indicates that interest rates can influence the company's dividend policy. When interest rates are low, companies tend to have more funds available for dividend payments to shareholders. Conversely, when interest rates are high, a company may choose to retain more profits for reinvestment purposes or pay higher interest on debt.

## 6. CONCLUSION

In light of the comprehensive examination and discussion of the research findings, the following conclusions can be articulated as follows:

1. **Impact on Investment Decisions:** The research reveals a weak and negative correlation between interest rates and company investment decisions. Fluctuations in interest rates is not significantly shape a company's approach to investments. A decline in interest rates prompts a more proactive stance, leading companies to initiate new investments or expand existing ones. This is attributed to the reduction in funding costs and the augmented availability of financial resources. Conversely, an upswing in interest rates tends to induce a more conservative outlook in company investment decisions (Hidayat, M, Bachtiar, N , Sjafrizal, Primayesa, 2021).

2. **Influence on Funding Decisions:** Interest rates exhibit a negative and weak influence on a company's funding decisions. Variations in interest rates play a pivotal role in determining how companies procure funds. During periods of low interest rates, companies are inclined to seek external funding sources, such as bank loans or bonds, to finance new projects. Conversely, in high-interest rate scenarios, companies may opt for internal funding sources or debt reduction strategies to mitigate funding costs (Hidayat, M, Bachtiar, N , Sjafrizal, Primayesa, 2021).

3. **Effect on Dividend Decisions**: The research findings indicate that interest rates do not wield a significant impact on a company's dividend decisions. Changes in interest rates demonstrate a relatively muted influence on the company's dividend policy. In periods of low interest rates, companies often find themselves with increased funds that can be allocated for shareholder dividends. However, during high-interest rate phases, companies may prioritize retaining profits for reinvestment purposes or directing higher dividends toward debt interest payments (Hidayat, M, Bachtiar, N , Sjafrizal, Primayesa, 2021). Based on the analysis, we find that this research finding is similar to the impact of interest rates on firm's financing policies of US industrial. So that, firms do not adjust their capital structures based on interest rates,

**Acknowledgments**

The authors gratefully acknowledge Scopus Camp 5 committee for helpful remarks on the previous version of the manuscript.